# A Locational Marginal Pricing Mechanism for Uncertainty Management Based on Improved Multi-Ellipsoidal Uncertainty Set

Zongzheng Zhao, Yixin Liu, Li Guo, Linquan Bai, and Chengshan Wang

*Abstract*—Large-scale integration of renewable energy sources (RES) brings huge challenges to the power system. A cost-effective reserve deployment and uncertainty pricing mechanism are critical to deal with the uncertainty and variability of RES. To this end, this paper proposes a novel locational marginal pricing mechanism in day-ahead market for managing uncertainties from RES. Firstly, an improved multi-ellipsoidal uncertainty set (IMEUS) considering the temporal correlation and conditional correlation of wind power forecast is formulated to better capture the uncertainty of wind power. The dimension of each ellipsoidal subset is optimized based on a comprehensive evaluation index to reduce the invalid region without large loss of modeling accuracy, so as to reduce the conservatism. Then, an IMEUS-based robust unit commitment (RUC) model and a robust economic dispatch (RED) model are established for the day-ahead market clearing. Both the reserve cost and ramping constraints are considered in the overall dispatch process. Furthermore, based on the Langrangian function of the RED model, a new locational marginal pricing mechanism is developed. The uncertainty locational marginal price (ULMP) is introduced to charge the RES for its uncertainties and reward the generators who provide reserve to mitigate uncertainties. The new pricing mechanism can provide effective price signals to incentivize the uncertainty management in the day-ahead market. Finally, the effectiveness of the proposed methods is verified via numerous simulations on the PJM 5-bus system and IEEE 118-bus system.

## I. INTRODUCTION

ALTHOUGH large-scale integration of renewable energy sources (RES) alleviates the environmental pressure, they bring great challenges to the power system [1]. The uncertainty and variability of RES need to be effectively managed in the market operation, calling for new pricing mechanisms for uncertainty management in the electricity markets.

The security-constrained economic dispatch (SCED) model is used in the current market-clearing to optimally dispatch the generators and provide price signals, i.e. the locational marginal price (LMP), to market participants. The LMP is composed of three components: energy price, congestion price and loss price. The traditional LMP does not account for uncertainties in the system. However, in today's electricity market with high penetrations of RES, it is essential to formulate a pricing mechanism considering uncertainty to guarantee the security and economy of system operation. Recently, some researchers have utilized SCED to design and derive the LMP mechanism for pricing system uncertainty. The SCED model in [2] is formulated as a two-stage stochastic programming problem, where the first stage clears the market and the second stage models the system operation under wind power uncertainty. A marginal pricing mechanism including pool energy prices and balancing energy prices is established. However, this mechanism only considers a single period market clearing, ignoring the impact of time coupling constraints such as ramp rate on market prices. Ref. [3] proposes a novel market framework to credit the generation and reserve and to charge the load and uncertainty within the SCED model in the day-ahead market. The units offer zero prices for their reserve products, and the marginal prices for clearing reserve in some periods are zero consequently. In addition, this paper considers the ramping constraints in both the dispatch and redispatch processes. Ref. [4] puts forward a stochastic market-clearing SCED model with an energy-only pricing scheme to yield LMPs. But it is unable to provide the pricing information for both the reserve and uncertainty. Ref. [5] derives an uncertainty contained LMP (U-LMP) which includes two new uncertainty components, i.e. transmission line overload price and generation violation price. Unlike traditional LMP, the U-LMP differs between different units and loads even at the same bus due to their uncertainty levels, which exacerbates the difficulty and complexity of market clearing.

On the other hand, the security-constrained unit commitment (SCUC) model is essential to determine the startup and shutdown status of conventional generating units before SCED calculation. Both robust SCUC (RSCUC) and robust SCED (RSCED) have been wildly used because they do not require detailed probability distribution of uncertain variables as well as complex calculation with a large number of scenarios. One of the key obstacles of the robust optimization (RO) methods is the uncertainty set modeling, which directly affects the

This work was supported in part by the National Key R&D Program of China (2020YFE0200400) and the National Nature Science Foundation of China (51907140).

Z. Zhao, Y. Liu (corresponding author), L. Guo, and C. Wang are with the Key Laboratory of Smart Grid of Ministry of Education, Tianjin University, Tianjin 300072, China (e-mail: triz425@tju.edu.cn; liuyixin@tju.edu.cn; liguo@tju.edu.cn; cswang@tju.edu.cn).

L. Bai is with the Department of Systems Engineering and Engineering Management, University of North Carolina at Charlotte, Charlotte, NC 28223, USA(e-mail: linquanbai@uncc.edu).

economy and robustness of the decision-making. Common uncertainty sets include box set, polyhedron set, and ellipsoidal set [6]. The box set utilizes upper and lower bounds to describe uncertainty, but ignores the correlation widely existing in RES generation and load demand [7]. As a result, the box set may contain numerous scenarios with extremely low occurrence probability, aggravating the conservativeness of RO model. The polyhedron set is able to capture the correlation, but still contains large invalid space, which cannot effectively reduce the conservativeness of RO model. Compared with the above two uncertainty sets, the ellipsoidal set can better preserve the correlation between variables, and has been proved to be effective in improving modeling accuracy and reducing conservatism. Ref. [8] proposes a novel RSCUC model where ellipsoidal uncertainty set (EUS) is adopted to well fit the correlated wind power and a novel criterion for budget value selection of the EUS is presented to reduce the conservatism. Ref. [9] presents a risk-averse two-stage optimization model. The first-stage problem minimizes the dispatch cost using the wind power forecast. The second-stage problem minimizes the expected redispatch cost in the worst-case where the minimum volume enclosing ellipsoid (MVEE) algorithm is used to construct the uncertainty set of correlated wind power. Ref. [10] proposes an adjustable RSCED model with wind power uncertainty. The MVEE is employed as a convex hull to solve the scenario-based RO model. However, [9] and [10] ignore the temporal correlation between wind power, which may lead to extreme climbing phenomenon of power units [11]. Furthermore, the temporal correlation of uncertain variables, e.g. wind power as well as its forecast error, are normally inversely proportional to their time interval distance. As a result, the high-dimensional ellipsoidal set, e.g. a 24-dimensional ellipsoidal set used in the conventional day-ahead SCUC and SCED problems, may contain a large number of uncertain variables with weak correlation, increasing the volume and hindering the conservatism reduction of the ellipsoidal set.

In view of the above shortcomings, this paper proposes a RO framework for day-ahead market-clearing based on an improved multi-ellipsoidal uncertainty set (IMEUS). A novel locational marginal pricing mechanism for pricing and managing uncertainties in the electricity market is built. In this work, the IMEUS is proposed to better characterize the uncertainties of wind power to reduce the conservatism of RO. Based on the proposed IMEUS, a RSCUC model and a RSCED model are established to optimize dispatch scheme for thermal units and generate price signals for energy and reserve respectively. The main contributions of this paper are summarized as follows:

1) An IMEUS modeling method that comprehensively considers the temporal correlation of the forecast errors and the conditional correlation between the forecast errors and the forecast values is proposed. Two indexes associated with the data coverage effectiveness and efficiency are formulated to realize dimension reduction and recombination of traditional high-dimensional ellipsoidal set. The proposed IMEUS can significantly reduce the conservatism of RO.

2) A day-ahead IMEUS-based RSCUC model and a RSCED model are established. The generation cost for reserve provision of units and the ramping constraints in the overall dispatch process are considered to guarantee the economy and safety operation of the units and power system. The optimized power output and reserve capacity of thermal units can meet the energy and uncertainty demand and avoid waste of flexible resources.

3) A novel locational marginal pricing mechanism is developed. The LMP is used for pricing energy. The uncertainty locational marginal price (ULMP) is introduced to charge the RES for its uncertainties and reward the generators who provide reserve to mitigate uncertainties. This pricing mechanism can provide effective price signals to guarantee the cost recovery of units and incentivize the uncertainty management.

The rest of this paper is organized as follows. Section II proposes the IMEUS model. Section III introduces the IMEUS-based RSCUC model. Section IV presents the new locational marginal pricing mechanism. Section V presents simulation results. Section VI concludes this paper with major findings.

## II. IMPROVED MULTI-ELLIPSOIDAL UNCERTAINTY SET

Without loss of generality, the uncertainty of wind power is addressed via the proposed IMEUS. The Gaussian copula, as an effective way to model the high-dimensional dependence structures [12], has been used to generate wind power samples considering the temporal correlation between the forecast error as well as the conditional correlation between the forecast error and the forecast value. Then the traditional high-dimensional ellipsoidal uncertainty set is decomposed into multiple low-dimensional subsets with an optimal dimension, which is reached by a comprehensive evaluation index, and finally the intersection of these subsets is taken as the IMEUS.

### A. Gaussian Copula-based Wind Power Sampling

Suppose $T$ is the number of time intervals in a day; $x_t$, $y_t$ and $e_t$ are the actual value, forecast value and forecast error of wind power at time $t$ respectively, $x_t=y_t+e_t$, $t=1,\ldots,T$. In Gaussian copula theory, the joint distribution of forecast error and forecast value can be written as [12]:

$$f_{e,y}(e,y)=\phi(z_{x,1},\ldots,z_{x,T},z_{y,1},\ldots,z_{y,T}|\boldsymbol{R}) \prod_{t=1}^{T} f_{x_t}(y_t+e_t)f_{y_t}(y_t) \quad (1)$$

where $z_{x,t}=\Phi_0^{-1}[F_{x,t}(y_t+e_t)]$; $z_{y,t}=\Phi_0^{-1}[F_{y,t}(y_t)]$; $f(\cdot)$ is the probability density function (PDF); $F(\cdot)$ is the cumulative distribution function (CDF); $\Phi_0^{-1}$ denotes the inverse of the CDF of the standard Gaussian distribution $\Phi_0(\cdot)$. $F_{x,t}(y_t+e_t)$ and $F_{y,t}(y_t)$ can be calculated from historical data. $\Phi_R(\cdot)$ is the standard Gaussian distribution with $\boldsymbol{R}$ as the covariance matrix. Let $\boldsymbol{z}=[\boldsymbol{z_x},\boldsymbol{z_y}]^T=[z_{x,1},\ldots,z_{x,T},z_{y,1},\ldots,z_{y,T}]^T$. $\boldsymbol{z}$ obeys standard Gaussian distribution, i.e. $\boldsymbol{z} \sim N(\boldsymbol{0},\boldsymbol{R})$. The $\boldsymbol{R}$ can be calculated from historical data, and its block matrix can be expressed as:

$$\boldsymbol{R}=\begin{bmatrix} \boldsymbol{R_{xx}} & \boldsymbol{R_{xy}} \\ \boldsymbol{R_{yx}} & \boldsymbol{R_{yy}} \end{bmatrix} \quad (2)$$

Given the day-ahead forecast value, the conditional distribution $\boldsymbol{z_x}|\boldsymbol{z_y} \sim N(\boldsymbol{\mu_{z_x|z_y}},\boldsymbol{R_{z_x|z_y}})$ can be obtained [13], where

$$\begin{cases} \boldsymbol{\mu_{z_x|z_y}}=\boldsymbol{R_{xy}}\boldsymbol{R_{yy}}^{-1}\boldsymbol{z_y} \\ \boldsymbol{R_{z_x|z_y}}=\boldsymbol{R_{xx}}-\boldsymbol{R_{xy}}\boldsymbol{R_{yy}}^{-1}\boldsymbol{R_{yx}} \end{cases} \quad (3)$$

The day-ahead wind power can be sampled based on the above results. The specific steps are as follows:

1) Obtain the historical data of wind power. Calculate $F_{x,t}(e_t+y_t)$, $F_{y,t}(y_t)$ and $\boldsymbol{R}$;

2) Calculate $z_{y,t}=\Phi_0^{-1}[F_{y,t}(y_t)]$ based on the latest day-ahead

forecast value of wind power $y_t$, $t=1,...,T$;

3) Calculate $\mu_{z_x|z_y}$ and $R_{z_x|z_y}$ in $z_x|z_y$ by (3);

4) Obtain samples of $z_x$ by sampling $z_x|z_y$ and then get the day-ahead wind power samples $x$ through $x_t = F_{x,t}^{-1}[\Phi_0(z_{x,t})]$.

### B. The IMEUS Model

The obtained wind power samples are used for the IMEUS modeling in this section. The expression of ellipsoidal uncertainty set with confidence degree $\alpha_c$ is:

$$(x-\mu_x)^T R_x^{-1}(x-\mu_x) \leq C_\alpha \quad (4)$$

where $\mu_x$ and $R_x$ are expectation vector and covariance matrix of samples $x$ respectively; $C_\alpha$ is a constant corresponding to $\alpha_c$.

$\mu_x$ and $R_x$ can be easily calculated from samples $x$. Take samples $x$ into (5) to obtain the CDF of $C$. Let $P(C \leq C_\alpha) = \alpha_c$, and then we can get $C_\alpha$ corresponding to $\alpha_c$.

$$C = (x-\mu_x)^T R_x^{-1}(x-\mu_x) \quad (5)$$

The conservatism of ellipsoidal uncertainty set is adjustable by changing $C_\alpha$. The larger $C_\alpha$ is, the more conservative the ellipsoidal uncertainty set will be.

The traditional method considers a 24-dimensional EUS to model the temporal correlation of wind power. However, weak correlation among distant periods and high dimensionality make EUS conservative. To this end, the integrity index and efficiency index are put forward for dimension reduction. The rolling modeling process is shown in Fig. 1. Firstly, the $T$-dimensional traditional ellipsoidal set is decomposed into $EN$ $OD$-dimensional ellipsoidal subsets chronologically. The $en$-th set $\Omega_{en}$ expressed as (6) is during time $[en, en+OD-1]$, where $1 \leq en \leq EN$, $2 \leq OD \leq T$, $EN=T-OD+1$. Then the intersection of these subsets are taken as the IMEUS $\Omega$, shown in (7).

$$\Omega_{en} = \{x | (x_{en}-\mu_{x,en})^T R_{x,en}^{-1}(x_{en}-\mu_{x,en}) \leq C_\alpha\} \quad (6)$$

$$\Omega = \bigcap_{en=1}^{EN} \Omega_{en} \quad (7)$$

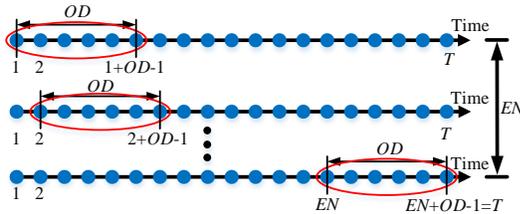

Fig. 1. Modeling process of the IMEUS.

The key of IMEUS modeling is to optimize $OD$ which is achieved by the integrity index and efficiency index. The integrity index $\zeta_{OD}$ is defined as the coverage ratio of the IMEUS $\Omega$ to the wind power data for $D$ days, which is used to ensure that more actual data is contained in the IMEUS. If the IMEUS $\Omega_d$ of the $d$-th day ($1 \leq d \leq D$) contains $N_d$ time intervals of actual wind power, $\zeta_{OD}$ can be expressed as:

$$\zeta_{OD} = (1/D) \sum_{d=1}^{D}(N_d/T) \quad (8)$$

The efficiency index $\eta_{OD}$ evaluates the volume of IMEUS $\Omega$ by referring to the box set $\Omega_{box}$ which is shown in (9). This index limits the volume of IMEUS to reduce the conservatism.

$$\Omega_{box} = \{x_t | x_t \in [x_t^{min}, x_t^{max}]\} \quad (9)$$

$N_{box,d}$ samples are generated in box set $\Omega_{box,d}$ of the $d$-th day, and the number of these samples in $\Omega_d$ is recorded as $N_{ell,d}$. The definition of $\eta_{OD}$ is as:

$$\eta_{OD} = 1 - (log_{10}(\tfrac{1}{D}\sum_{d=1}^{D} N_{ell,d}) / log_{10}(\tfrac{1}{D}\sum_{d=1}^{D} N_{box,d})) \quad (10)$$

Taking into account the two indexes comprehensively, the comprehensive index $I_{OD}$ is defined as:

$$I_{OD} = k \times \zeta_{OD} + (1-k) \times \eta_{OD}, \; k \in [0,1] \quad (11)$$

Since there are few potential solutions, the exhaustive method is used to optimize $OD$. The pseudocode for optimizing $OD$ is demonstrated in Table I.

TABLE I
PSEUDOCODE FOR OPTIMIZING $OD$

| Require: Wind power samples obtained in Section II-A. |
|---|
| 1: For $OD=2$: $T$, do |
| 2:     For $d=1$:$D$, do |
| 3:        Calculate the IMEUS $\Omega_{en,d}$. |
| 4:        Calculate $N_d$, $N_{ell,d}$ and $N_{box,d}$. |
| 5:     End for |
| 6:     Calculate $\zeta_{OD}$, $\eta_{OD}$ and $I_{OD}$. |
| 7: End for |
| 8: Select the $OD$ with maximum value of $I_{OD}$ as the solution. |

The IMEUS is further rewritten as second-order cone form as (12) for wind power, and the uncertainty budget is introduced to adjust the conservative degree flexibly.

$$P^w := \begin{cases} P_t^w \mid \|C_{\alpha,en}^{-1/2} L_{en}(P_{en}^w - \mu_{en})\|_2 \leq 1 \\ P_t^w \geq B_{w,t} P_{f,t}^w \\ \sum_{t=1}^T B_{w,t} \geq \Gamma^w \\ t=1,...,T,\; en=1,...,EN \end{cases} \quad (12)$$

where $P_{en}^w = [P_{en}^w,...,P_{en+OD-1}^w]^T$ is the wind power during time $[en, en+OD-1]$. $\mu_{en}$ and $C_{\alpha,en}$ are parameters in (4). $L_{en}$ can be obtained by Cholesky decomposition $R_{en}^{-1} = L_{en}^T L_{en}$, where $R_{en}^{-1}$ is the inverse of the covariance matrix $R_{en}$. $P_{f,t}^w$ is the day-ahead forecast value at time $t$. $B_{w,t}$ is a binary variable related to the realization of wind power. $\Gamma^w$ is the uncertainty budget, which is integer valued between 0 and $T$, denoting the minimum number of periods when wind power is taken as the forecast value. The solution is more conservative when $\Gamma^w$ is smaller.

### III. ROBUST SECURITY-CONSTRAINED UNIT COMMITMENT

#### A. RSCUC Model

The RSCUC model optimizes the dispatch scheme of energy and reserve with the lowest cost in the worst-case scenario, and provides unit commitment status and the worst-case uncertainty realization for RSCED calculation. It contains basic dispatch and redispatch processes in this paper. In basic dispatch process, units provide energy according to the forecast value of load and wind power. In the redispatch process, unit reserve is optimized to cope with the deviation between the forecast value and the worst-case realization of load and wind power. Considering the fact that the forecast accuracy of load demand is higher than that of wind power, the uncertainty of load is addressed only by box sets. The RSCUC model is formulated as:

$$\min_{u,v,I,P^g,\Delta P^g} \sum_{t=1}^T \sum_{i=1}^{N_G} (C_i^g(P_{i,t}^g + \Delta P_{i,t}^g) + C_i^{SU} u_{i,t} + C_i^{SD} v_{i,t}) \quad (13)$$

subject to

$$\sum_{i=1}^{N_G} P_{i,t}^g + \sum_{j=1}^{N_W} P_{j,f,t}^w = \sum_{m=1}^{N_M} P_{m,f,t}^d, \forall t \; (\lambda_t^b) \quad (14)$$

$$-P_{i,t}^g \geq -I_{i,t} P_{i,\max}^g, \forall i,t \; (\overline{\beta}_{i,t}^b) \quad (15)$$

$$P_{i,t}^g \geq I_{i,t} P_{i,\min}^g, \forall i,t \; (\underline{\beta}_{i,t}^b) \quad (16)$$

$$P_{i,t-1}^g - P_{i,t}^g \geq -r_i^{ru}(1-u_{i,t}) - r_i^{su} u_{i,t}, \forall i,t \; (\overline{\alpha}_{i,t}^b) \quad (17)$$

$$P_{i,t}^g - P_{i,t-1}^g \geq -r_i^{rd}(1-v_{i,t}) - r_i^{sd} v_{i,t}, \forall i,t \; (\underline{\alpha}_{i,t}^b) \quad (18)$$

$$-\sum_{m=1}^{N_M} GSF_{l,m} P_{m,t}^{inj} \geq -F_l, \forall l,t \; (\overline{\eta}_{l,t}^b) \quad (19)$$

$$\sum_{m=1}^{N_M} GSF_{l,m} P_{m,t}^{inj} \geq -F_l, \forall l,t \; (\underline{\eta}_{l,t}^b) \quad (20)$$

$$\sum_{i=1}^{N_G} \Delta P_{i,t}^g = \sum_{m=1}^{N_M} \Delta P_{m,t}^d - \sum_{j=1}^{N_W} \Delta P_{j,t}^w \; (\lambda_t^r) \quad (21)$$

$$-P_{i,t}^g - \Delta P_{i,t}^g \geq -I_{i,t}^g P_{i,max}^g, \forall i,t \; (\overline{\beta}_{i,t}^r) \quad (22)$$

$$P_{i,t}^g + \Delta P_{i,t}^g \geq I_{i,t}^g P_{i,min}^g, \forall i,t \; (\underline{\beta}_{i,t}^r) \quad (23)$$

$$-\Delta P_{i,t}^g \geq -r_i^{ru}(1-u_{i,t}) - r_i^{su} u_{i,t}, \forall i,t \; (\overline{\alpha}_{i,t}^{r1}) \quad (24)$$

$$\Delta P_{i,t}^g \geq -r_i^{rd}(1-v_{i,t}) - r_i^{sd} v_{i,t}, \forall i,t \; (\underline{\alpha}_{i,t}^{r1}) \quad (25)$$

$$P_{i,t-1}^g + \Delta P_{i,t-1}^g - P_{i,t}^g - \Delta P_{i,t}^g \geq -r_i^{ru}(1-u_{i,t}) - r_i^{su} u_{i,t}, \forall i,t \; (\overline{\alpha}_{i,t}^{r2}) \quad (26)$$

$$P_{i,t}^g + \Delta P_{i,t}^g - P_{i,t-1}^g - \Delta P_{i,t-1}^g \geq -r_i^{rd}(1-v_{i,t}) - r_i^{sd} v_{i,t}, \forall i,t \; (\underline{\alpha}_{i,t}^{r2}) \quad (27)$$

$$-\sum_{m=1}^{N_M} GSF_{l,m}(P_{m,t}^{inj} + \Delta P_{m,t}^{inj}) \geq -F_l, \forall l,t \; (\overline{\eta}_{l,t}^r) \quad (28)$$

$$\sum_{m=1}^{N_M} GSF_{l,m}(P_{m,t}^{inj} + \Delta P_{m,t}^{inj}) \geq -F_l, \forall l,t \; (\underline{\eta}_{l,t}^r) \quad (29)$$

$$-\sum_{q=t-UT_i^g+1}^{t} u_{i,q} \geq -I_{i,t}, \forall i,t \in [UT_i^g, T] \quad (30)$$

$$-\sum_{q=t-DT_i^g+1}^{t} v_{i,q} \geq I_{i,t}-1, \forall i,t \in [DT_i^g, T] \quad (31)$$

$$u_{i,t} - v_{i,t} = I_{i,t} - I_{i,t-1}, u_{i,t}, v_{i,t}, I_{i,t} \in \{0,1\}, \forall i,t \quad (32)$$

$$-u_{i,t} - v_{i,t} \geq -1, \forall i,t \quad (33)$$

where $i$, $j$, $m$, and $l$ denote index for conventional generating units (except RES), wind farms, buses, and transmission lines; $N_G$, $N_W$, and $N_M$ are the number of conventional units, wind farms, and loads respectively; $P_{i,t}^g$ is the energy output (MW) of unit $i$ at time $t$ in the basic dispatch process; $\Delta P_{i,t}^g$ is the unit reserve (MW) in the redispatch process; $C_i^g(\cdot)$ denotes the operational cost of unit $i$; $C_i^{SU}$ and $C_i^{SD}$ are startup cost and shutdown cost of unit $i$ respectively; $u_{i,t}$ is the startup variable (1 for startup, 0 otherwise); $v_{i,t}$ is the shutdown variable (1 for shutdown, 0 otherwise); $I_{i,t}$ is the unit commitment binary variable (1 for online, 0 otherwise); $P_{j,f,t}^w$ and $P_{m,f,t}^d$ are day-ahead forecast power (MW) of wind farm $j$ and load at bus $m$; $GSF_{l,m}$ is the generation shift factor of bus $m$ to line $l$; $F_l$ is the maximum transmission flow (MW) of line $l$. $P_{i,min}^g$ and $P_{i,max}^g$ denote lower and upper output limits (MW) of unit $i$ respectively; $I_{i,t}^g$ represents the on/off status of unit $i$ at time $t$; $UT_i^g$ and $DT_i^g$ are minimum up time and minimum down time (h) respectively; $r_i^{ru}$ and $r_i^{rd}$ are maximum ramp-up/down rate (MW/h) of unit $i$ respectively; $r_i^{su}$ and $r_i^{sd}$ are startup and shutdown ramp rate (MW/h) respectively; $\Delta P_{m,t}^d$ and $\Delta P_{j,t}^w$ denote the difference between forecast value and the worst-case realization of load ($P_{m,t}^d$) and wind power ($P_{j,t}^w$); $P_{m,t}^{inj}$ and $\Delta P_{m,t}^{inj}$ are power injection (MW) in basic dispatch process and incremental power injection in redispatch process. $\Delta P_{m,t}^d$, $\Delta P_{j,t}^w$, $P_{m,t}^{inj}$, and $\Delta P_{m,t}^{inj}$ can be expressed as:

$$\Delta P_{m,t}^d = P_{m,t}^d - P_{m,f,t}^d \quad (34)$$
$$\Delta P_{j,t}^w = P_{j,t}^w - P_{j,f,t}^w \quad (35)$$
$$P_{m,t}^{inj} := \sum_{i \in G(m)} P_{i,t}^g + \sum_{j \in W(m)} P_{j,f,t}^w - P_{m,f,t}^d \quad (36)$$
$$\Delta P_{m,t}^{inj} := \sum_{i \in G(m)} \Delta P_{i,t}^g + \sum_{j \in W(m)} \Delta P_{j,t}^w - \Delta P_{m,t}^d \quad (37)$$

where $G(m)$ and $W(m)$ denote the set of wind farms and loads at bus $m$ respectively.

The objective function (13) minimizes the total operation costs over the entire dispatch horizon. Constraints (14)-(20), (21)-(29), and (30)-(33) relate to the basic dispatch, redispatch and unit on/off status respectively. (14) and (21) ensure power balance at each bus. (15)-(16) and (22)-(23) enforce generation limits of units. (17)-(18) and (24)-(27) denote ramping constraints. (19)-(20) and (28)-(29) enforce the line capacity. (30) and (31) impose the minimum up/down time constraints. (32) describes the relationship between the unit commitment, startup and shutdown variables. (33) ensures that startup and shutdown of one unit cannot occur at the same time.

The proposed RSCUC model includes three parts of ramping constraints. The first two parts, namely (17)-(18) and (24)-(25), present requirements for the energy output and reserve capacity of units, respectively. These constraints has been applied in the RSCUC model [3]. Different from the previous works, this paper adds constraints (26)-(27) as the third part to ensure that the aggregate of energy and reserve of units does not violate ramping limits. In practical situation, even if one unit meets the ramping constraints in terms of both energy output and reserve capacity, the sum of energy and reserve may exceed ramping limits in some extreme scenarios when the reserve capacity is fully deployed by the independent system operator (ISO). The energy output and reserve capacity of units are mutually restricted. (26)-(27) guarantee the safe operation of the units.

In addition, the objective function (13) includes the reserve cost of units, i.e. the term $C_i^g(\Delta P_{i,t}^g)$, which guarantees the economy of the reserve deployment and the overall RSCUC model. Furthermore, the price signals derived from the reserve cost-contained RSCED model (described in section IV) can ensure that the reserve revenue covers the reserve cost of units.

The uncertainty of wind power and load is considered in the RSCUC model. The IMEUS of wind power is shown in (12). The box set of load is expressed as

$$\boldsymbol{P}^d := \begin{cases} P_t^d \mid P_t^d = P_{f,t}^d + B_{d,t} \Delta P_t^d \\ \sum_{t=1}^T B_{d,t} \leq \Gamma^d \end{cases} \quad (38)$$

where $B_{d,t}$ is a binary variable. $\Gamma^d$ is the uncertainty budget, which is similar to $\Gamma^w$ shown in (12).

Equations (12)-(33) and (38) can be modeled as a two-stage "min-max-min" RO model shown as:

$$\begin{cases} \min_{\boldsymbol{x}} \boldsymbol{c}^T \boldsymbol{x} + \max_{\boldsymbol{u} \in U} \min_{\boldsymbol{y} \in \Omega(\boldsymbol{x},\boldsymbol{u})} \boldsymbol{b}^T \boldsymbol{y} \\ \text{s.t.} \quad \boldsymbol{Gx} + \boldsymbol{Ey} + \boldsymbol{Mu} \geq \boldsymbol{h} \\ \left\| C_{a,en}^{1/2} L_{en}(\boldsymbol{P}_{en}^w - \boldsymbol{\mu}_{en}) \right\|_2 \leq 1, \; en=1,...,EN \end{cases} \quad (39)$$

where $\boldsymbol{c}$ and $\boldsymbol{b}$ denote the cost coefficient of generation offer, startup and shutdown of units shown in (13); $\boldsymbol{x}$, $\boldsymbol{y}$, and $\boldsymbol{u}$ are decision variables summarized as (40); $\Omega(\boldsymbol{x},\boldsymbol{u})$ denotes the feasible region of $\boldsymbol{y}$ for a given set of $(\boldsymbol{x},\boldsymbol{u})$.

$$\boldsymbol{x}=[\boldsymbol{I}_i, \boldsymbol{u}_i, \boldsymbol{v}_i]^T, \boldsymbol{y}=[\boldsymbol{P}_i^g, \Delta \boldsymbol{P}_i^g]^T, \boldsymbol{u}=[\boldsymbol{P}_m^d, \boldsymbol{P}_j^w]^T \quad (40)$$

where $\boldsymbol{I}_i$, $\boldsymbol{u}_i$, and $\boldsymbol{v}_i$ denote the vectors of unit commitment variables, startup variables, and shutdown variables of unit $i$ during $[0, T]$ respectively; $\boldsymbol{P}_i^g$ and $\Delta \boldsymbol{P}_i^g$ are the vectors of energy and reserve of unit $i$ during $[0, T]$ respectively.

### B. Solution Method

The column-and-constraint generation (CCG) algorithm [14] is utilized to solve the RSCUC model. The optimal solution is obtained by decomposing the original problem (39) into the master problem (MP) and subproblem (SP) and solving them alternatively. The MP can be expressed as

$$\begin{cases} \min_{x} \ c^T x + \alpha \\ \text{s.t.} \ \alpha \geq b^T y_k \\ Gx + Ey_k + Mu_k^* \geq h \\ \|C_{a,en}^{1/2} L_{en} (P_{en,k}^{w*} - \mu_{en})\|_2 \leq 1 \end{cases} \quad (41)$$

where $k$ is the iteration number; $y_k$ is the SP solution in the $k$-th iteration; $u_k^*$ is the worst-case scenario realization of wind power and load in the $k$-th iteration.

The SP can be expressed as

$$\max_{u \in U} \min_{y \in \Omega(x,u)} b^T y \quad (42)$$

The above "max-min" bi-level program can be transformed into a single level "max" model according to strong duality theory. The bilinear terms included in the objective function can be linearized by the big-M approach. As a result, the SP can be reformulated as a mixed integer second order cone programming (MISOCP) problem. The detailed solution method and process can be found in [14].

After the above transformation, the CCG algorithm-based solution steps are as follows:

1) Give a set of $u$ as the initial worst-case scenario, set the lower bound of the objective function (39) $LB=-\infty$, and the upper bound $UB=+\infty$, $k=1$;

2) Solve the MP (41), obtain the solution $(x_k^*, \alpha_k^*, y_k^*)$, update the lower bound as $LB = c^T x_k^* + \alpha_k^*$;

3) Fix $x_k^*$ and solve the SP (42), obtain the objective function $f_k^*(x_k^*)$ and the worst-case scenario $u_{k+1}^*$, update the upper bound as $UB = \min\{UB, c^T x_k^* + f_k^*(x_k^*)\}$;

4) Set optimality tolerance $\varepsilon$. If $UB-LB \leq \varepsilon$, the iteration is terminated and return the optimal solution $x_k^*$, $y_k^*$, and $u_{k+1}^*$; otherwise create $y_{k+1}$ and add following constraints to the MP

$$\begin{cases} \alpha \geq b^T y_{k+1} \\ Gx + Ey_{k+1} + Mu_{k+1}^* \geq h \\ \|C_{a,en}^{1/2} L_{en} (P_{en,k+1}^{w*} - \mu_{en})\|_2 \leq 1 \end{cases} \quad (43)$$

Update $k=k+1$ and go to step 2).

## IV. LOCATIONAL MARGINAL PRICING MECHANISM

The ISO should deploy reserve appropriately and form a pricing mechanism in a cost-effective manner to guide power entities to improve system operation efficiency while ensuring their own profits. This section first expounds the RSCED model, and then derives a novel locational marginal pricing mechanism.

### A. RSCED Model

Fixing the unit commitment status and robust uncertainty scenario realization (i.e. $I_{i,t}$, $u_{i,t}$, $v_{i,t}$, $P_{m,t}^d$ and $P_{j,t}^w$) obtained in the RSCUC model, the RSCED model can be formulated as:

$$\min_{P^g, \Delta P^g} \sum_{t=1}^{T} \sum_{i=1}^{N_G} C_i^g (P_{i,t}^g + \Delta P_{i,t}^g) \quad (44)$$

s.t. (14)-(29)

The variables in brackets of (14)-(29) are the corresponding dual variables. The RSCED is a linear programming model which can be easily solved by optimization software.

### B. Pricing Mechanism and Market Clearing Mechanism

Lagrangian function $L(P, \Delta P, \lambda, \beta, \alpha, \eta)$ [4] is a by-product of the RSCED model. According to the definition, the LMP at bus $m$ can be derived as:

$$\begin{aligned} LMP_{m,t} &= \partial L(P, \Delta P, \lambda, \beta, \alpha, \eta) / \partial P_{m,t}^d \\ &= \lambda_t^b - \sum_l GSF_{l,m}(\overline{\eta}_{l,t}^b - \underline{\eta}_{l,t}^b) - \sum_l GSF_{l,m}(\overline{\eta}_{l,t}^r - \underline{\eta}_{l,t}^r) \end{aligned} \quad (45)$$

The ULMP is defined as the marginal cost corresponding to the unit increment of forecast deviation of net load at bus $m$, which can be derived as:

$$\begin{aligned} ULMP_{m,t} &= \partial L(P, \Delta P, \lambda, \beta, \alpha, \eta) / \partial (\Delta P_{m,t}^d - \sum_{j \in W(m)} \Delta P_{j,t}^w) \\ &= \lambda_t^r - \sum_l GSF_{l,m}(\overline{\eta}_{l,t}^r - \underline{\eta}_{l,t}^r) \end{aligned} \quad (46)$$

For the unit $i$ at bus $m$, LMP and ULMP can also be derived from KKT condition [4]:

$$\begin{aligned} LMP_{m,t} = & \partial C_i^g(P_{i,t}^g + \Delta P_{i,t}^g) / \partial P_{i,t}^g + \overline{\beta}_{i,t}^b - \underline{\beta}_{i,t}^b + \overline{\alpha}_{i,t}^b - \underline{\alpha}_{i,t}^b - \overline{\alpha}_{i,t+1}^b \\ & + \underline{\alpha}_{i,t+1}^b + \overline{\beta}_{i,t}^r - \underline{\beta}_{i,t}^r + \overline{\alpha}_{i,t}^{r2} - \underline{\alpha}_{i,t}^{r2} - \overline{\alpha}_{i,t+1}^{r2} + \underline{\alpha}_{i,t+1}^{r2} \end{aligned} \quad (47)$$

$$\begin{aligned} ULMP_{m,t} = & \partial C_i^g(P_{i,t}^g + \Delta P_{i,t}^g) / \partial \Delta P_{i,t}^g + \overline{\beta}_{i,t}^r - \underline{\beta}_{i,t}^r + \overline{\alpha}_{i,t}^{r1} - \underline{\alpha}_{i,t}^{r1} \\ & + \overline{\alpha}_{i,t}^{r2} - \underline{\alpha}_{i,t}^{r2} - \overline{\alpha}_{i,t+1}^{r2} + \underline{\alpha}_{i,t+1}^{r2} \end{aligned} \quad (48)$$

It can be seen from (45)-(48) that the marginal prices are affected by various factors such as line capacity and unit capacity [15] [16], which belong to non-time coupling factors. On the other hand, the time coupling factors, mainly the unit ramping constraints, have been proved to cause high marginal prices possibly [17]-[19]. Different from the previous works, this paper proves that ramping constraints may lead to low prices (lower than the generation cost of units). The dual variables $\overline{\alpha}_{i,t}^b$, $\underline{\alpha}_{i,t}^b$, $\overline{\alpha}_{i,t}^{r1}$, $\underline{\alpha}_{i,t}^{r1}$, $\overline{\alpha}_{i,t}^{r2}$, and $\underline{\alpha}_{i,t}^{r2}$ in (47) and (48) correspond to the ramping constraints (17)-(18) and (24)-(27). When one of these constraints is binding, its dual variable is greater than 0. The following two cases may cause low prices: 1) The ramp up limits are triggered at $t+1$, so $\overline{\alpha}_{i,t+1}^b > 0$ or $\overline{\alpha}_{i,t+1}^{r2} > 0$. 2) One unit reaches its ramp down limits at time $t$, thus $\underline{\alpha}_{i,t}^b > 0$, $\underline{\alpha}_{i,t}^{r1} > 0$, or $\underline{\alpha}_{i,t}^{r2} > 0$. Both cases can decrease LMP and ULMP at time $t$, which may happen at the beginning of ramp up stage or the end of ramp down stage.

Constraints (21)-(29) denote the reserve supply process to cope with the system uncertainties, and the corresponding dual variables are included in the marginal price expressions (45)-(48). Thus LMP and ULMP indicate not only the energy price and congestion price, but also the uncertainty level of the power system, although LMP is only a partial differential of the deterministic load. The two marginal prices reflect the demand intensity for system resources such as energy and unit climbing capacity, which are influenced by the uncertainties.

Based on the definition, LMP is used to price energy, ULMP is used to price uncertainty and reserve. The day-ahead market clearing scheme is as follows:

*1) Energy Income and Expenditure:*

The energy income of unit $i$ at bus $m$ is the product of the energy output and LMP, i.e. $LMP_{m,t} P_{i,t}^g$. The energy expenditure of the load and the income of the wind farm $j$ at bus $m$ is the product of their power forecast and LMP, i.e. $LMP_{m,t} P_{m,f,t}^d$ and $LMP_{m,t} P_{j,f,t}^w$ respectively.

*2) Expenditure of Uncertainty Source:*

The expenditures for uncertainties of the load and the wind farm $j$ at bus $m$ are $ULMP_{m,t} \Delta P_{m,t}^d$ and $-ULMP_{m,t} \Delta P_{j,t}^w$, which is determined by the maximum forecast deviation and the ULMP.

*3) Income for Generation Reserve:*

The income for reserve provision of unit $i$ at bus $m$ is the product of its reserve capacity and ULMP, i.e. $ULMP_{m,t} \Delta P_{i,t}^g$. The term $\partial C_i^g(P_{i,t}^g + \Delta P_{i,t}^g)/\partial \Delta P_{i,t}^g$ of ULMP in (48) related to the reserve cost in the RSCED model ensures that units can receive enough revenue commensurate with their reserve cost.

The uncertainties increase the system cost and the expenditure of uncertainty sources, which are ultimately embodied in the changes of LMP and ULMP. The proposed pricing mechanism can stimulate uncertainty sources to improve forecast accuracy and provide effective price signals to incentivize the uncertainty management in the day-ahead market.

## V. CASE STUDY

The effectiveness of the proposed method is verified by the PJM 5-bus system and IEEE 118-bus system in this section. All the simulations have been implemented in MATLAB 2017a with YALMIP interface and CPLEX12.9.0 in a computational environment with Intel(R) Core(TM) i7-10700F CPU running at 2.9 GHz with 16 GB RAM.

### A. The IMEUS Method

A group of hourly data in 2019 is taken from a 400MW wind farm in China for research. The temporal correlation coefficient of forecast error is shown in Fig. 2(a), where the colors indicate the correlation strength and the abscissa/ordinate represent time interval. The forecast error shows obvious temporal correlation which is inversely proportional to time interval distance. This correlation is related to the wind power fluctuation and affect the ramp rate of units [20]. The conditional correlation between forecast error and forecast value is presented in Fig. 2(b). The forecast error increases and the error distribution becomes more dispersed as the forecast value becomes larger. The formulated Gaussian Copula function establishes the relationship between forecast error and forecast value, so that the samples of the possible realization of wind power can be updated based on the latest day-ahead forecast power. The temporal correlation and conditional correlation are important features of wind power, which need to be fully considered in uncertainty set modeling.

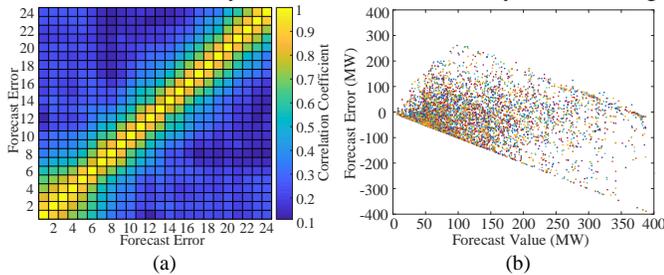

Fig. 2. (a) Temporal correlation coefficient of forecast error of wind power. (b) Conditional correlation between forecast error and forecast value of wind power.

The relationship between the dimension *OD* of an ellipsoidal subset and the evaluation indexes is illustrated in Fig. 3 when $k$ and the confidence degree are set to 0.3 and 90% respectively. The integrity index increases monotonously with the increase of *OD*, whereas the efficiency index decreases gradually after $OD \geq 5$ as more invalid region is contained in the IMEUS. The comprehensive index performs best when $OD = 6$, which is selected as the optimal solution for *OD*.

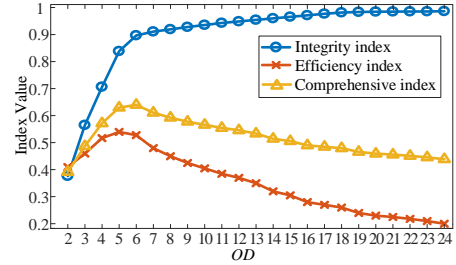

Fig. 3. Relation between the dimension of a subset and the evaluation indexes.

To verify the advantage of the IMEUS, three uncertainty set modeling methods are compared. Method 1 utilizes the box set that adds error interval on both sides of the forecast value in each period [6]. Method 2 uses the traditional high-dimensional ellipsoidal set [6]. Method 3 applies the proposed IMEUS model. The confidence levels are all set to 90%. The data among January to November is used to train the model, and the data in the last month is applied to test the performance, which is summarized in Table II. The coverage ratio denotes the total proportion of the actual data in the uncertainty sets, and the average width is the average difference between the upper and lower bounds of the uncertainty sets.

TABLE II
PERFORMANCE OF DIFFERENT UNCERTAINTY SET MODELING METHODS

| Index | Method 1 | Method 2 | Method 3 |
|---|---|---|---|
| Coverage ratio | 90.11% | 99.85% | 90.07% |
| Average width | 95.84MW | 130.87MW | 80.53MW |

Although the traditional ellipsoidal set performs better in terms of actual data coverage ratio, it contains too much region and is therefore, too conservatism. Whereas the IMEUS has the narrowest interval on the basic of satisfactory performance in terms of coverage ratio, thus can reduce the conservatism. The coverage ratio of the box set is similar to that of IMEUS, but the conservatism of the IMEUS is still lower. The reason is that the IMEUS updates the set via the latest day-ahead forecast data to reflect the temporal correlation, whereas the box set with fixed error interval each day shows poor adaptability. Figure 4 demonstrates the set realization of three methods on a specified day, where the above conclusions can also been drawn. These results indicate that the IMEUS can effectively reduce the conservatism on the premise of ensuring the modeling accuracy.

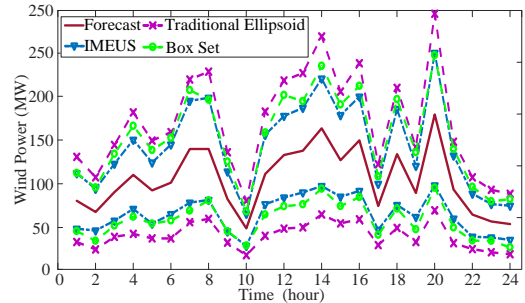

Fig. 4. Wind power uncertainty sets of the three modeling methods.

### B. The IMEUS-based RSCUC Model and RSCED Model

The PJM 5-bus system is shown in Fig. 5. The units originally located at bus A are combined and unit parameters are shown in Table III [21]. The forecast total load is distributed to the loads at bus B, C, and D with a ratio of 3:3:4. It is assumed that there is no forecast error for the load at bus D, and the maximum forecast error for the loads at bus B and C are 10% and 5% of their forecast value respectively. The wind farm mentioned above is located at bus D.

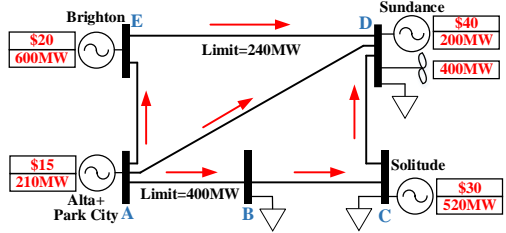

Fig. 5. The PJM 5-bus system.

TABLE III
UNIT PARAMETERS

| Unit Name | $r^{ru}/r^{su}/r^{rd}/r^{sd}$ (MW/h) | $C^{su}$ ($) | $C^{sd}$ ($) | $UT^g$ (h) | $DT^g$ (h) |
|---|---|---|---|---|---|
| A(Alta+Park City) | 25 | 360 | 40 | 4 | 3 |
| C(Solitude) | 60 | 500 | 80 | 4 | 4 |
| D(Sundance) | 25 | 300 | 50 | 2 | 2 |
| E(Brighton) | 80 | 550 | 90 | 3 | 3 |

The proposed RSCUC model is implemented with the three uncertainty set modeling methods for wind power mentioned in V-A, respectively. The load is modeled by box set in all cases. The uncertain budgets are all set as 24. The daily average performances of the 31-days (in November) simulations are listed in Table IV. Thanks to the improved conservatism, the extreme wind power fluctuation scenarios with low probability of occurrence are eliminated in the IMEUS. As a result, the ramp-up and ramp-down rate levels of wind power are lowest in the RSCUC model based on the IMEUS (Method 3). In addition, the IMEUS-based RSCUC can also reduce the reserve demand, and the RSCUC cost is significantly reduced by 1.86% as compared with Method 1 and 7.53% with Method 2. To sum up, the IMEUS reduces the RO conservatism and the demand for flexible resources, and improves the economy and reliability.

TABLE IV
RSCUC RESULTS UNDER DIFFERENT UNCERTAINTY MODELING METHODS

| Daily average | Maximum ramp-up of wind power (MW) | Maximum ramp-down of wind power (MW) | Reserve capacity (MW) | RSCUC cost ($) |
|---|---|---|---|---|
| Method 1 | 82.17 | 87.00 | 1487.31 | 257112 |
| Method 2 | 91.00 | 111.08 | 1915.49 | 272882 |
| Method 3 | 76.00 | 85.10 | 1328.62 | 252342 |

Fix the worst-case scenario realization obtained by the IMEUS-based RSCUC model and then compare the following three different RSCUC/RSCED models:

Model 1: the proposed RSCUC/RSCED models in this paper.

Model 2: the RSCUC/RSCED models proposed in [3]. Different from Model 1, the overall ramping constraints ((26) and (27)) and the unit reserve cost ($C_i^g(\Delta P_{i,t}^g)$ in (13)) are not considered in this model. The derived ULMP does not include the cost term $\partial C_i^g(P_{i,t}^g+\Delta P_{i,t}^G)/\partial \Delta P_{i,t}^g$ in (48).

Model 3: all the ramping constraints, i.e. (17), (18), and (24)-(27) are excluded, other aspects are the same as Model 1.

Figure 6 indicates the dispatching scheme. Each period includes three histograms, corresponding to the results of three models respectively. It can be seen that the load is mainly supported by units A and E due to their low generation cost. In Model 1, units A, C, D, and E generate 4859.02MWh, 554.52MWh, 3.00MWh and 6345.46MWh respectively. In Model 2, units A, C, D, and E generate 4921.00MWh, 282.15MWh, 14.25MWh and 6544.57MWh respectively.

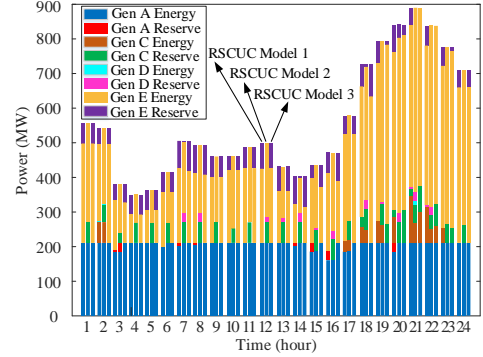

Fig. 6. Dispatching results of the three RSCUC models.

In Model 1, the residual capacity of unit A with the cheapest offer is limited after providing energy. Therefore, the reserve capacity is mainly provided by unit E with large residual capacity and low offer price. Units A, C, D and E provide 99.94MW, 64.12MW, 12.9MW, and 1442.65MW reserve respectively. In Model 2, the reserve cost is not considered, thus units C and D with higher cost supply a great quantity of reserve capacity. Specifically, units A, C, D and E provide 25.00MW, 1310.89MW, 231.53MW and 52.20MW reserve respectively.

The above results show that the proportion of energy provided by each unit in Model 1 and Model 2 is close. However, the two models differ greatly in reserve provision. In Model 1, unit E with lower cost supplies 89.07% reserve. In Model 2, unit C with higher cost provides 80.94% reserve. As a result, the energy costs of the two models are close, whereas the reserve cost and total cost of Model 1 are significantly less than those of Model 2. The energy, reserve, and total costs of Model 1/Model 2 are 216550.14$/213742.05$, 32791.89$/50006.87$, 249342.03$/263748.92$, respectively.

The proposed RSCUC model tightens the ramping constraints. Taking hour 17-18 as an example, the energy output of unit E increases from 307MW to 376.87MW, the reserve capacity increases from 54.64MW to 64.78MW, and the total output will increase from 361.64MW to 441.64MW if the reserve is fully dispatched in Model 1. Both the energy output and reserve capacity do not violate the ramping constraints, yet their sum just triggers the limit of 80MW. In Model 2, however, the total output of the energy and reserve increases from 305.00MW to 393.56MW, exceeding the ramping limit. A more serious situation occurs in Model 3, where both the energy output (from 313.00MW to 423.00MW) and the overall output (from 367.64MW to 516.56MW) violate the constraints. Accordingly, the proposed RSCUC/RSCED model ensures the safe operation of units by considering ramping constraints comprehensively.

### C. Pricing Mechanism and Day-ahead Market Clearing

The LMP and ULMP derived from the three RSCED models mentioned in section V-B are shown in Fig. 7 and Fig. 8. In Model 1, the prices are high during heavy load periods in hour 18-22. The LMP or ULMP is different at different buses in each period of hour 20-23 due to the network congestion on line D-E. In addition, there are some differences between LMP and ULMP. For example, LMP is 15$/MWh higher than ULMP in hour 2 (40$/MWh vs 25$/MWh). It can be explained by (47)-

(48), and the marginal prices at bus A is taken as an example firstly. $\partial C_i^g(P_{i,t}^g+\Delta P_{i,t}^g)/\partial P_{i,t}^g$ and $\partial C_i^g(P_{i,t}^g+\Delta P_{i,t}^g)/\partial \Delta P_{i,t}^g$ of unit A (at bus A) are both equal to its generation cost, i.e. 15$/MWh. Besides, unit A reaches the output limit in hour 2 and ramp down limit from hour 2 to hour 3, consequently the results of $\underline{\alpha}_{A,3}^b$ and $\overline{\beta}_{A,2}^r$ are 15$/MWh and 10$/MWh respectively. Then the marginal prices at bus A are obtained by introducing these nonzero terms into (47)-(48). The prices of other buses can be derived in the same way. Alternatively, since there is no network congestion, each bus has the same marginal prices, which can be obtained directly according to the prices of bus A.

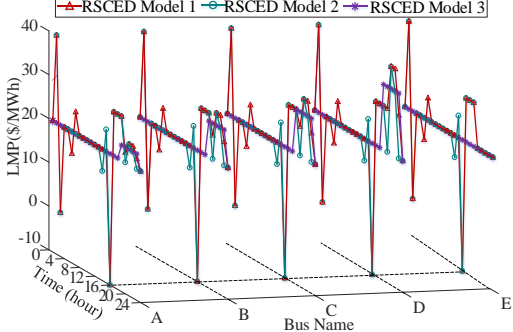

Fig. 7. LMP derived from the three RSCED models.

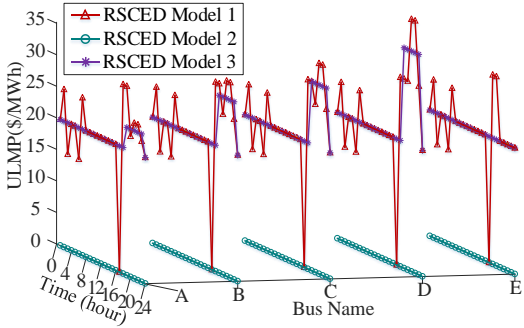

Fig. 8. ULMP derived from the three RSCED models.

The three RSCED models produce different prices due to the different consideration of reserve cost and ramping constraints. The LMP of Model 1 and Model 2 deviate in hour 6-7, 14-15, 20, and 22-23. For instance, the LMP of Model 1 is 5$/MWh higher than that of Model 2 in hour 7 (25$/MWh vs 20$/MWh). The cheap units A and E in Model 1 provide all the reserve in this period, and yet the expensive units C and D in Model 2 also provide reserve due to the lack of reserve cost. The full use of the output capacity of unit E makes the sum of energy and reserve reach the ramping limit (from 214.41MW in hour 6 to 294.41MW in hour 7) in Model 1. As a result, the $\overline{\alpha}_{i,t}^{r2}$ of unit E equals to 5$/MWh in Model 1 with constraint (26) binding, whereas there is no such term in Model 2. Thus LMP at bus E in Model 1 and Model 2 are 25$/MWh and 20$/MWh according to (47) as $\partial C_i^g(P_{i,t}^g+\Delta P_{i,t}^g)/\partial P_{i,t}^g$=20$/MWh for unit E. The LMP of other buses is the same as that of bus E due to no congestion.

The ULMP in Model 1 almost matches the reserve cost of units. However, the ULMP in Model 2 are all equal to 0$/MWh. The reason is that the reserve cost term $\partial C_i^g(P_{i,t}^g+\Delta P_{i,t}^g)/\partial \Delta P_{i,t}^g$ is considered in the ULMP expression in Model 1, but is ignored in Model 2. The unit income/profit from reserve provision in Model 1 and Model 2 is 33677.20$/885.30$ and 0$/-50006.87$, respectively. The proposed model provides enough revenue for reserve provision of units, whereas Model 2 cannot cover the reserve cost, resulting in profit loss. Model 3 considers the reserve cost but ignores the ramping constraints. LMP and ULMP are 20$/MWh in hour 1-18 and 24, when there is no network congestion and unit E is the marginal unit. In hour 19-23, unit C is activated as the marginal unit due to the congestion of line D-E, and the price at each bus goes up consequently. Model 3 guarantees the unit profits, but the lack of ramping constraints increases the operation risk of the system. In a conclusion, the proposed pricing mechanism can generate effective price signals to reflect the actual safety operation of the system and ensure the cost recovery of units simultaneously.

Fig. 7 and Fig. 8 reveal that low electricity prices (less than the offer prices of units at corresponding buses) occur in hour 3, 6, 16, and 17 when units trigger ramping limits in Model 1. There are no low prices when ramping limits are not considered in Model 3. Low prices lead to profit loss of units in that period. Hour 15-19 coupled with ramping constraints are selected for analysis. The profits except unit D (all zero values) are presented in Table V (at the left of the solidus "/"). Units A and E suffer losses due to low prices in hour 16 and 17, but their total profits (6250$ and 4800$ respectively) are positive in hour 15-19. Units A and E have an incentive to reduce output to minimize profit loss. Suppose that unit A reduce 1MW energy in hour 16 and unit E reduce 1MW reserve in hour 17. To meet the system constraints, other units follow to adjust energy and reserve in hour 15-19. After the adjustment, the profit change of units is shown in Table V (at the right of the solidus "/").

TABLE V
UNIT PROFITS AND PROFIT CHANGE IN HOUR 15-19 ($)

| Energy profit | 15h | 16h | 17h | 18h | 19h |
|---|---|---|---|---|---|
| Unit A | 925/-5 | -4000/25 | 2775/-15 | 3150/-15 | 3150/-167 |
| Unit C | 0/0 | 0/-80 | 0/0 | 0/0 | 0/0 |
| Unit E | 0/0 | -6810/30 | 3070/-10 | 3769/91 | 4569/91 |
| Reserve profit | 15h | 16h | 17h | 18h | 19h |
| Unit A | 125/0 | 125/0 | 0/0 | 0/0 | 0/167 |
| Unit C | 0/0 | 0/0 | 0/-30 | 0/0 | 0/0 |
| Unit E | 0/0 | 0/0 | -1093/20 | 648/-111 | 648/-111 |

Unit A in hour 16 and unit E in hour 17 reduce profit loss by cut down energy and reserve respectively. However, due to the output adjustment in the rest hours, the profits of units A and C decrease 10$ and 110$ respectively, and the profit of unit E keeps unchanged. The system costs before/after adjustment are 56703.06$/56823.15$ in hour 15-19. Thus, the units should not deviate from the system optimal plan in terms of both the system cost and unit revenue. Another verification is that low prices may appear at the beginning of ramp up stage (hour 6 and 16) and the end of ramp down stage (hour 3 and 17). From the previous discussion, it can be seen that low prices at time $t$ can be caused by a unit reaching the ramp down limit from time $t$-1 to $t$ or the ramp up limit from time $t$ to $t$+1. In order to gain enough revenue by exporting more energy and reserve in high price periods (e.g. time $t$-1 and $t$+1), the units constrained by the ramping constraints still provide a certain amount of energy and reserve in low price periods (e.g. time $t$), even though they may bear some loss. When these periods are considered as a whole, each unit can get a considerable positive profit.

Table VI shows day-ahead market clearing results. Because the energy output provided by all the units is far more than the reserve capacity, the unit profit mainly comes from energy supply, accounting for 97.62% of the total profit. Loads and wind farms should pay tariff for their uncertainties. The reserve expenditure of wind farm with strong uncertainty accounts for 33.54% of its energy income. Therefore, it is urgent to improve forecast accuracy to improve regulation ability.

TABLE VI
DAY-AHEAD MARKET CLEARING RESULTS ($)

| Economic index | Energy | Reserve | Total |
|---|---|---|---|
| Load expenditure | 335618.28 | 14512.86 | 350131.13 |
| Wind income | 55625.40 | -22047.83 | 33577.56 |
| Unit Profit | 36419.64 | 885.30 | 37304.94 |

*D. IEEE 118-Bus System*

The performance of the proposed method is also tested in the IEEE 118-bus system, which includes 118 buses, 54 traditional thermal units and 186 lines. The detailed parameters are given in [3]. This system contains 10 uncertain loads at bus (11, 15, 49, 54, 56, 59, 60, 62, 80, and 90). Five wind farms are located at bus (10, 26, 49, 65, and 80). The load and wind power are modeled by box set and the IMEUS model respectively.

The forecast deviation of uncertain load is set as 0.2, 0.25 and 0.3 of the forecast value respectively, and the simulation results are demonstrated in Table VII. It can be seen that the aggravation of load uncertainties raises the LMP and ULMP of the whole network. The energy and reserve expenditures of the load increase accordingly. Moreover, the energy income and reserve expenditure of the wind farms also increase, which is affected by the prices change of the system. In a conclusion, the serious system uncertainties make electricity prices go up, and influences the income and expenditure of each participant, not only the uncertainty sources themselves.

TABLE VII
DAY-AHEAD CALCULATION RESULTS

| Load deviation | System cost ($) | LMP/ ULMP average ($/MWh) | Energy payment (load) ($) | Reserve payment (load) ($) | Energy income (wind) ($) | Reserve payment (wind) ($) |
|---|---|---|---|---|---|---|
| 0.2 | $2.01\times 10^6$ | 20.18 | $2.71\times 10^6$ | $1.58\times 10^5$ | $2.10\times 10^5$ | $9.22\times 10^4$ |
| 0.25 | $2.05\times 10^6$ | 20.45 | $2.75\times 10^6$ | $2.01\times 10^5$ | $2.14\times 10^5$ | $9.36\times 10^4$ |
| 0.3 | $2.10\times 10^6$ | 21.94 | $2.97\times 10^6$ | $2.60\times 10^5$ | $2.28\times 10^5$ | $9.99\times 10^4$ |

## VI. CONCLUSION

This paper proposes a novel locational marginal pricing mechanism in day-ahead market for managing uncertainties. The IMEUS is proposed to better characterize the uncertainties of wind power to reduce the conservatism of RO problems. The IMEUS-based RSCUC model and RSCED model are presented to optimize dispatch scheme for thermal units and generate price signals for energy, reserve and uncertainty respectively.

The following conclusions can be drawn from the simulations: 1) The IMEUS model can reduce the conservativeness of uncertainty set and the RSCUC/ RSCED models, and improve the economy and reliability of system operation. 2) The proposed RSCUC/RSCED models strengthen the ramping constraints to realize the safe operation of units. The proposed ULMP ensures the cost recovery of units, thus can motivate the participants to provide reserve services. The system operation cost is reduced by the proposed methods. 3) The unit ramping constraint is one of the reasons for the low marginal prices. Low prices may cause profit loss of units in that period, but the unit profits of the whole day can be guaranteed.